# Light Curve Modelling and Evolutionary Status of the Short Period Binary 1SWASP J092328.76+435044

M. M. Elkhateeb[1,2], M. I. Nouh[1,2], R. Michel[3], A. Haroon[2,4] and E. Elkholy[1,2]

[1]Department of Physics, College of Science, Northern Border University, Arar, Saudi Arabia.

Email: abdo_nouh@hotmail.com

[2]Department of Astronomy, National Research Institute of Astronomy and Geophysics, Helwan, Cairo17211, Egypt

[3]Instituto de Astronomía, Universidad Nacional Autónoma de México, Apartado Postal 877, Ensenada, Baja California, C.P. 22830 México

[4]Department of Astronomical Science, Faculty of Science, King Abdulaziz University, Jeddah, Saudi Arabia

## 1. Abstract

Light curve modeling for the newly discovered super contact low-mass W UMa system 1SWASP J092328.76+435044 was carried out by using a new BVR complete light curves. A spotted model was applied to treat the asymmetry of the light curves. The output model was obtained by means of Wilson-Devinney code, which reveals that the massive component is hotter than the less massive one with about $\Delta T \approx 40^0 K$. A total of 6 new times of minima were estimated. The evolutionary state of the system components was investigated based on the estimated physical parameters.

## 2. Introduction

Overcontact binaries defined as a short-period limit at about 0.22 days (Rucinski, 1992; 2007), their period – color relation reveals that the shorter period corresponds to the later type components with fainter brightness (Jiang et al. 2015). This means that it's difficult to detect overcontact binaries with the period of 0.22 days. Light curve analysis of W UMa overcontact binaries gives information about their stellar properties, The eclipsing binary candidate 1SWASP J092328.76+435044 (here after we shall refer to it as J0923 for simplicity) was announced as a new discovered variable star together with several tens of thousands by early February 2012 through the super WASP project which surveying bright (V ≈ 8-15 magnitude) stars across almost the whole sky since 2004 (Lohr et al. 2012).

The system J0923 is classified as a short period W UMa star (p = $0.^d2349$, $V_{max}$ = 13.03 mag.) with primary and secondary depth = $0.^m04$ and $0.^m03$ respectively (Norton et al, 2011). The significant period increase of about +1.90±0.95 sec/year was calculated for the system by Lohr et al. 2012, 2013.

In this paper, we present a continuation of a program started earlier to study some newly discovered eclipsing binaries by Elkhateeb et al., 2014a, b,c, 2015a, b, and 2016. A detailed light curve analysis for the candidate J0923 based on our new BVR observations and evolution state using the preliminary physical parameters have been presented.

## 3. Observations

A total of 570 CCD photometric observations represented complete light curves in BVR bands for the system J0923 were carried out on 2016 March 21 using the 84 cm telescope with SI-Te4 CCD of San Pedro Martir Observatory (Mexico). The period of the system J0923 was announced as $0.^d21013$ and $0.^d23487$ by Norton et al. (2011), and Lohr et al. (2012) respectively. As is clear, there is a difference between the two values by $0.^d02474$ (36 min), we tried to use both values of period in calculating the photometric phases of the observed BVR light curves, but the estimated values of phase show a shift in both the two minima, which means that the used period should be modified.

A new period was estimated, P = $0.^d2376$ which seems to be longer than calculated before and reveals photometric phases with zero phase shift as shown in Fig. 1. The miscellaneous values of the period for the system J0923 since its discovery in 2010 may give an impression about period instability, which requires continuous observations and set of minima to follow and estimated any period variation if present. The individual phases corresponding to each observed data were calculated using the first ephemeris adopted using our observations as:

$$\text{Min I} = 2457468.7316 + 0.2376 * E \qquad (1)$$

Differential photometry was performed with respect to 2MASSJ09240120+4350218 and 2MASSJ09232921+4355371 as a comparison and check stars respectively. The corresponding phased BVR light curves are plotted in Fig. 1, which shows a typical EW-type with an obvious asymmetry, the original data are listed in Table 1.

A total of 6 new times of minima (3 primary and 3 secondaries, each in one of the B, V, or R filters) were obtained using the Minima V2.3 Package (Nelson, 2006) based on the Kwee and Van Worden (1956) method; they are listed in Table 2. A first ephemeris was determined from the present minima.

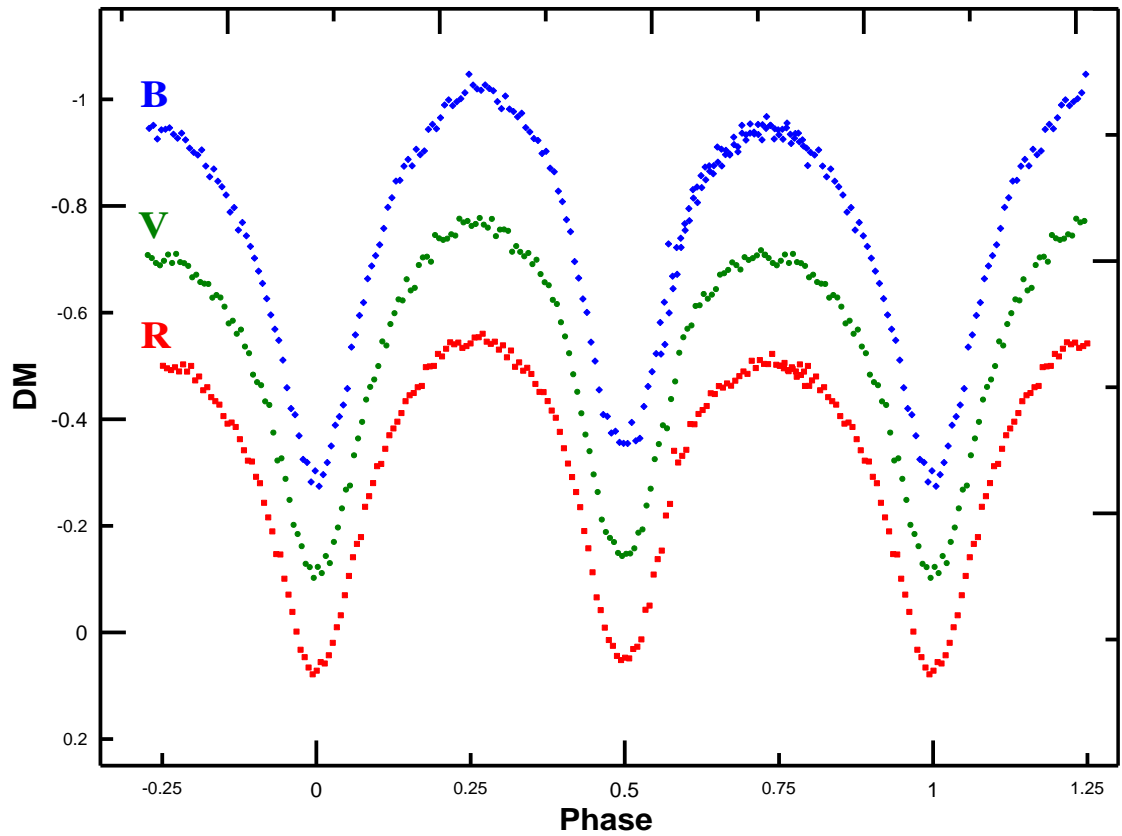

Fig. 1 *BVR* light curves of J0923.

Table 1 BVR observational data of the eclipsing binary J0923

| JD | Phase | ΔB | JD | Phase | ΔV | JD | Phase | ΔR |
|---|---|---|---|---|---|---|---|---|
| 2457468.6733 | 0.7545 | -0.6951 | 2457468.6727 | 0.7519 | -0.6984 | 2457468.6722 | 0.7502 | -0.7515 |
| 2457468.6748 | 0.7610 | -0.6977 | 2457468.6743 | 0.7586 | -0.7101 | 2457468.6700 | 0.7569 | -0.7467 |
| 2457468.6764 | 0.7675 | -0.6856 | 2457468.6758 | 0.7651 | -0.6943 | 2457468.6754 | 0.7634 | -0.7434 |
| 2457468.6779 | 0.7740 | -0.6783 | 2457468.6774 | 0.7717 | -0.7114 | 2457468.6769 | 0.7699 | -0.7485 |
| 2457468.6795 | 0.7805 | -0.6878 | 2457468.6789 | 0.7782 | -0.6959 | 2457468.6785 | 0.7764 | -0.7407 |
| 2457468.6810 | 0.7870 | -0.6753 | 2457468.6805 | 0.7847 | -0.6939 | 2457468.6800 | 0.7830 | -0.7541 |
| 2457468.6826 | 0.7936 | -0.6596 | 2457468.6820 | 0.7912 | -0.6888 | 2457468.6816 | 0.7895 | -0.7406 |
| 2457468.6841 | 0.8001 | -0.6518 | 2457468.6836 | 0.7978 | -0.6674 | 2457468.6831 | 0.7960 | -0.7511 |
| 2457468.6857 | 0.8067 | -0.6469 | 2457468.6851 | 0.8043 | -0.6723 | 2457468.6847 | 0.8025 | -0.7239 |
| 2457468.6872 | 0.8132 | -0.6564 | 2457468.6867 | 0.8108 | -0.6585 | 2457468.6862 | 0.8091 | -0.7312 |
| 2457468.6888 | 0.8197 | -0.6260 | 2457468.6882 | 0.8174 | -0.6555 | 2457468.6878 | 0.8156 | -0.7055 |
| 2457468.6903 | 0.8262 | -0.6060 | 2457468.6898 | 0.8239 | -0.6549 | 2457468.6894 | 0.8222 | -0.7112 |
| 2457468.6919 | 0.8328 | -0.6206 | 2457468.6913 | 0.8305 | -0.6292 | 2457468.6909 | 0.8287 | -0.6925 |
| 2457468.6934 | 0.8394 | -0.5976 | 2457468.6929 | 0.8370 | -0.6342 | 2457468.6925 | 0.8353 | -0.6850 |
| 2457468.6950 | 0.8459 | -0.5873 | 2457468.6944 | 0.8436 | -0.6295 | 2457468.6940 | 0.8418 | -0.6785 |
| 2457468.6966 | 0.8527 | -0.5719 | 2457468.6960 | 0.8501 | -0.6123 | 2457468.6956 | 0.8483 | -0.6569 |
| 2457468.6982 | 0.8592 | -0.5397 | 2457468.6976 | 0.8569 | -0.5809 | 2457468.6972 | 0.8551 | -0.6429 |
| 2457468.6997 | 0.8658 | -0.5482 | 2457468.6992 | 0.8634 | -0.5860 | 2457468.6987 | 0.8617 | -0.6450 |
| 2457468.7013 | 0.8725 | -0.5061 | 2457468.7007 | 0.8700 | -0.5615 | 2457468.7003 | 0.8682 | -0.6366 |
| 2457468.7029 | 0.8791 | -0.5203 | 2457468.7023 | 0.8767 | -0.5693 | 2457468.7019 | 0.8750 | -0.6136 |
| 2457468.7044 | 0.8856 | -0.4951 | 2457468.7039 | 0.8833 | -0.5375 | 2457468.7035 | 0.8815 | -0.5929 |
| 2457468.7060 | 0.8922 | -0.4753 | 2457468.7054 | 0.8898 | -0.5250 | 2457468.7050 | 0.8881 | -0.5735 |
| 2457468.7075 | 0.8987 | -0.4533 | 2457468.707 | 0.8964 | -0.4848 | 2457468.7066 | 0.8946 | -0.5717 |
| 2457468.7091 | 0.9053 | -0.4289 | 2457468.7085 | 0.9029 | -0.4707 | 2457468.7081 | 0.9012 | -0.5432 |
| 2457468.7107 | 0.9118 | -0.4060 | 2457468.7101 | 0.9095 | -0.4650 | 2457468.7097 | 0.9077 | -0.5311 |
| 2457468.7123 | 0.9186 | -0.3773 | 2457468.7117 | 0.916 | -0.4346 | 2457468.7112 | 0.9143 | -0.4943 |
| 2457468.7138 | 0.9251 | -0.3467 | 2457468.7133 | 0.9228 | -0.4279 | 2457468.7128 | 0.9210 | -0.4669 |
| 2457468.7154 | 0.9317 | -0.3201 | 2457468.7148 | 0.9293 | -0.3761 | 2457468.7144 | 0.9276 | -0.4406 |
| 2457468.7169 | 0.9382 | -0.2993 | 2457468.7164 | 0.9359 | -0.3237 | 2457468.7159 | 0.9341 | -0.3981 |
| 2457468.7185 | 0.9448 | -0.2621 | 2457468.7179 | 0.9424 | -0.3278 | 2457468.7175 | 0.9407 | -0.3974 |
| 2457468.7200 | 0.9513 | -0.2103 | 2457468.7195 | 0.949 | -0.2893 | 2457468.7191 | 0.9472 | -0.3520 |
| 2457468.7216 | 0.9579 | -0.1711 | 2457468.721 | 0.9555 | -0.2498 | 2457468.7206 | 0.9538 | -0.3220 |
| 2457468.7232 | 0.9644 | -0.1595 | 2457468.7226 | 0.9621 | -0.2030 | 2457468.7222 | 0.9603 | -0.2897 |
| 2457468.7247 | 0.9710 | -0.1201 | 2457468.7242 | 0.9686 | -0.1858 | 2457468.7237 | 0.9669 | -0.2527 |
| 2457468.7263 | 0.9775 | -0.0759 | 2457468.7257 | 0.9752 | -0.1629 | 2457468.7253 | 0.9734 | -0.2186 |
| 2457468.7278 | 0.9840 | -0.0702 | 2457468.7273 | 0.9817 | -0.1299 | 2457468.7268 | 0.9800 | -0.2046 |
| 2457468.7294 | 0.9905 | -0.0335 | 2457468.7288 | 0.9882 | -0.1237 | 2457468.7284 | 0.9865 | -0.1854 |
| 2457468.7309 | 0.9971 | -0.0542 | 2457468.7304 | 0.9948 | -0.1036 | 2457468.7299 | 0.9930 | -0.1725 |
| 2457468.7325 | 0.0036 | -0.0253 | 2457468.7319 | 0.0013 | -0.1241 | 2457468.7315 | 0.9995 | -0.1794 |
| 2457468.7340 | 0.0101 | -0.0472 | 2457468.7334 | 0.0078 | -0.1125 | 2457468.7330 | 0.0060 | -0.1954 |
| 2457468.7356 | 0.0167 | -0.0698 | 2457468.735 | 0.0143 | -0.1445 | 2457468.7346 | 0.0126 | -0.1928 |
| 2457468.7371 | 0.0232 | -0.1010 | 2457468.7366 | 0.0209 | -0.1312 | 2457468.7361 | 0.0191 | -0.2083 |
| 2457468.7387 | 0.0297 | -0.1400 | 2457468.7381 | 0.0274 | -0.1710 | 2457468.7377 | 0.0256 | -0.2318 |
| 2457468.7402 | 0.0362 | -0.1561 | 2457468.7397 | 0.0339 | -0.1980 | 2457468.7392 | 0.0321 | -0.2609 |
| 2457468.7418 | 0.0428 | -0.1778 | 2457468.7412 | 0.0405 | -0.2339 | 2457468.7408 | 0.0387 | -0.2833 |
| 2457468.7433 | 0.0493 | -0.2085 | 2457468.7428 | 0.047 | -0.2691 | 2457468.7423 | 0.0452 | -0.3209 |
| 2457468.7449 | 0.0559 | -0.2864 | 2457468.7443 | 0.0535 | -0.2767 | 2457468.7439 | 0.0517 | -0.3571 |
| 2457468.7464 | 0.0624 | -0.3093 | 2457468.7459 | 0.0601 | -0.3335 | 2457468.7455 | 0.0583 | -0.3921 |
| 2457468.7480 | 0.0690 | -0.3459 | 2457468.7474 | 0.0666 | -0.3651 | 2457468.7470 | 0.0648 | -0.4172 |
| 2457468.7495 | 0.0755 | -0.3700 | 2457468.749 | 0.0732 | -0.3962 | 2457468.7486 | 0.0714 | -0.4301 |
| 2457468.7511 | 0.0821 | -0.4147 | 2457468.7505 | 0.0797 | -0.4380 | 2457468.7501 | 0.0779 | -0.4869 |
| 2457468.7527 | 0.0886 | -0.4386 | 2457468.7521 | 0.0863 | -0.4618 | 2457468.7517 | 0.0845 | -0.5066 |
| 2457468.7542 | 0.0952 | -0.4580 | 2457468.7537 | 0.0928 | -0.4757 | 2457468.7532 | 0.0910 | -0.5313 |
| 2457468.7558 | 0.1018 | -0.4783 | 2457468.7552 | 0.0994 | -0.5009 | 2457468.7548 | 0.0976 | -0.5631 |
| 2457468.757 | 0.1083 | -0.5091 | 2457468.7568 | 0.106 | -0.5471 | 2457468.7564 | 0.1042 | -0.5672 |
| 2457468.7589 | 0.1149 | -0.5487 | 2457468.7583 | 0.1125 | -0.5395 | 2457468.7579 | 0.1108 | -0.5957 |
| 2457468.7605 | 0.1214 | -0.5667 | 2457468.7599 | 0.1191 | -0.5797 | 2457468.7595 | 0.1173 | -0.6210 |
| 2457468.7620 | 0.1280 | -0.5979 | 2457468.7615 | 0.1256 | -0.6001 | 2457468.7610 | 0.1239 | -0.6339 |
| 2457468.7636 | 0.1346 | -0.5993 | 2457468.763 | 0.1322 | -0.6260 | 2457468.7626 | 0.1304 | -0.6466 |
| 2457468.7651 | 0.1411 | -0.6258 | 2457468.7646 | 0.1388 | -0.6240 | 2457468.7642 | 0.1370 | -0.6627 |
| 2457468.7667 | 0.1477 | -0.6387 | 2457468.7661 | 0.1454 | -0.6637 | 2457468.7657 | 0.1435 | -0.6850 |
| 2457468.7683 | 0.1543 | -0.6265 | 2457468.7677 | 0.1519 | -0.6428 | 2457468.7673 | 0.1502 | -0.6960 |
| 2457468.7698 | 0.1608 | -0.6576 | 2457468.7693 | 0.1585 | -0.6476 | 2457468.7688 | 0.1567 | -0.7004 |
| 2457468.7714 | 0.1674 | -0.6474 | 2457468.7708 | 0.1651 | -0.6904 | 2457468.7704 | 0.1633 | -0.7113 |
| 2457468.7729 | 0.1740 | -0.6545 | 2457468.7724 | 0.1717 | -0.7047 | 2457468.7720 | 0.1699 | -0.7136 |
| 2457468.7745 | 0.1806 | -0.6950 | 2457468.774 | 0.1783 | -0.7061 | 2457468.7735 | 0.1765 | -0.7489 |
| 2457468.7761 | 0.1872 | -0.7043 | 2457468.7755 | 0.1848 | -0.6967 | 2457468.7751 | 0.1830 | -0.7503 |
| 2457468.7776 | 0.1938 | -0.6960 | 2457468.7771 | 0.1915 | -0.7467 | 2457468.7767 | 0.1896 | -0.7512 |
| 2457468.7792 | 0.2003 | -0.7167 | 2457468.7786 | 0.198 | -0.7406 | 2457468.7782 | 0.1962 | -0.7736 |
| 2457468.7808 | 0.2069 | -0.7407 | 2457468.7802 | 0.2046 | -0.7376 | 2457468.7798 | 0.2028 | -0.7689 |
| 2457468.7823 | 0.2134 | -0.7504 | 2457468.7818 | 0.2111 | -0.7399 | 2457468.7813 | 0.2094 | -0.7834 |
| 2457468.7839 | 0.2201 | -0.7394 | 2457468.7833 | 0.2177 | -0.7481 | 2457468.7829 | 0.2159 | -0.7954 |
| 2457468.7855 | 0.2266 | -0.7468 | 2457468.7849 | 0.2243 | -0.7457 | 2457468.7845 | 0.2225 | -0.7917 |
| 2457468.7870 | 0.2332 | -0.7525 | 2457468.7865 | 0.2309 | -0.7774 | 2457468.7860 | 0.2291 | -0.7949 |
| 2457468.7886 | 0.2398 | -0.7636 | 2457468.788 | 0.2375 | -0.7704 | 2457468.7876 | 0.2357 | -0.7851 |
| 2457468.7902 | 0.2464 | -0.7984 | 2457468.7896 | 0.2441 | -0.7731 | 2457468.7892 | 0.2423 | -0.7884 |
| 2457468.7917 | 0.2530 | -0.7780 | 2457468.7912 | 0.2506 | -0.7638 | 2457468.7907 | 0.2489 | -0.7933 |
| 2457468.7933 | 0.2596 | -0.7709 | 2457468.7927 | 0.2573 | -0.7675 | 2457468.7923 | 0.2555 | -0.8044 |
| 2457468.7948 | 0.2661 | -0.7682 | 2457468.7943 | 0.2638 | -0.7787 | 2457468.7939 | 0.2621 | -0.8050 |
| 2457468.7964 | 0.2727 | -0.7782 | 2457468.7959 | 0.2704 | -0.7662 | 2457468.7954 | 0.2686 | -0.8115 |
| 2457468.7980 | 0.2793 | -0.7717 | 2457468.7974 | 0.2769 | -0.7606 | 2457468.7970 | 0.2752 | -0.7967 |
| 2457468.7995 | 0.2859 | -0.7674 | 2457468.799 | 0.2835 | -0.7770 | 2457468.7985 | 0.2817 | -0.7925 |
| 2457468.8011 | 0.2924 | -0.7471 | 2457468.8005 | 0.2901 | -0.7446 | 2457468.8001 | 0.2883 | -0.7967 |
| 2457468.8027 | 0.2990 | -0.7338 | 2457468.8021 | 0.2967 | -0.7569 | 2457468.8017 | 0.2949 | -0.7816 |
| 2457468.8042 | 0.3056 | -0.7573 | 2457468.8037 | 0.3032 | -0.7572 | 2457468.8032 | 0.3015 | -0.7898 |
| 2457468.8058 | 0.3121 | -0.7330 | 2457468.8052 | 0.3098 | -0.7546 | 2457468.8048 | 0.3080 | -0.7672 |
| 2457468.8073 | 0.3187 | -0.7283 | 2457468.8068 | 0.3164 | -0.7154 | 2457468.8064 | 0.3146 | -0.7800 |
| 2457468.8089 | 0.3253 | -0.7183 | 2457468.8083 | 0.323 | -0.7260 | 2457468.8079 | 0.3212 | -0.7496 |
| 2457468.8105 | 0.3319 | -0.7253 | 2457468.8099 | 0.3295 | -0.7151 | 2457468.8095 | 0.3278 | -0.7579 |
| 2457468.8120 | 0.3384 | -0.6983 | 2457468.8115 | 0.3361 | -0.7068 | 2457468.8110 | 0.3343 | -0.7423 |
| 2457468.8136 | 0.3450 | -0.6902 | 2457468.813 | 0.3426 | -0.7130 | 2457468.8126 | 0.3409 | -0.7477 |
| 2457468.8152 | 0.3516 | -0.6776 | 2457468.8146 | 0.3493 | -0.6921 | 2457468.8142 | 0.3475 | -0.7360 |
| 2457468.8167 | 0.3583 | -0.6740 | 2457468.8162 | 0.3559 | -0.7004 | 2457468.8157 | 0.3541 | -0.7173 |
| 2457468.8183 | 0.3648 | -0.6498 | 2457468.8177 | 0.3625 | -0.6734 | 2457468.8173 | 0.3607 | -0.7025 |
| 2457468.8199 | 0.3715 | -0.6538 | 2457468.8193 | 0.3691 | -0.6581 | 2457468.8189 | 0.3673 | -0.7019 |
| 2457468.8214 | 0.3780 | -0.6220 | 2457468.8209 | 0.3757 | -0.6525 | 2457468.8204 | 0.3739 | -0.6852 |
| 2457468.8230 | 0.3847 | -0.6155 | 2457468.8224 | 0.3823 | -0.6251 | 2457468.8220 | 0.3805 | -0.6674 |
| 2457468.8246 | 0.3913 | -0.5790 | 2457468.824 | 0.3889 | -0.6173 | 2457468.8236 | 0.3871 | -0.6540 |
| 2457468.8261 | 0.3979 | -0.5594 | 2457468.8256 | 0.3955 | -0.5831 | 2457468.8251 | 0.3937 | -0.6274 |
| 2457468.8277 | 0.4045 | -0.5255 | 2457468.8272 | 0.4022 | -0.5565 | 2457468.8267 | 0.4003 | -0.5966 |
| 2457468.8293 | 0.4111 | -0.5030 | 2457468.8287 | 0.4087 | -0.5250 | 2457468.8283 | 0.4070 | -0.5681 |
| 2457468.8309 | 0.4177 | -0.4470 | 2457468.8303 | 0.4154 | -0.4855 | 2457468.8299 | 0.4136 | -0.5427 |
| 2457468.8324 | 0.4244 | -0.4159 | 2457468.8319 | 0.422 | -0.4526 | 2457468.8314 | 0.4202 | -0.5146 |
| 2457468.8340 | 0.4310 | -0.3769 | 2457468.8334 | 0.4286 | -0.4123 | 2457468.8330 | 0.4268 | -0.4862 |
| 2457468.8356 | 0.4376 | -0.3516 | 2457468.835 | 0.4352 | -0.3724 | 2457468.8346 | 0.4334 | -0.4414 |
| 2457468.8371 | 0.4442 | -0.2956 | 2457468.8366 | 0.4419 | -0.3412 | 2457468.8362 | 0.4400 | -0.4089 |
| 2457468.8387 | 0.4509 | -0.2602 | 2457468.8382 | 0.4484 | -0.2976 | 2457468.8377 | 0.4467 | -0.3641 |
| 2457468.8403 | 0.4574 | -0.2061 | 2457468.8397 | 0.4551 | -0.2646 | 2457468.8393 | 0.4533 | -0.3170 |
| 2457468.8419 | 0.4641 | -0.1599 | 2457468.8413 | 0.4617 | -0.2129 | 2457468.8409 | 0.4599 | -0.2929 |
| 2457468.8434 | 0.4707 | -0.1565 | 2457468.8429 | 0.4683 | -0.1896 | 2457468.8425 | 0.4665 | -0.2601 |
| 2457468.8450 | 0.4773 | -0.1253 | 2457468.8444 | 0.4749 | -0.1785 | 2457468.8440 | 0.4732 | -0.2368 |
| 2457468.8466 | 0.4839 | -0.1285 | 2457468.846 | 0.4815 | -0.1707 | 2457468.8456 | 0.4797 | -0.2260 |
| 2457468.8481 | 0.4905 | -0.1078 | 2457468.8476 | 0.4881 | -0.1503 | 2457468.8472 | 0.4863 | -0.2071 |
| 2457468.8497 | 0.4971 | -0.1059 | 2457468.8491 | 0.4947 | -0.1446 | 2457468.8487 | 0.4929 | -0.1993 |
| 2457468.8513 | 0.5037 | -0.1056 | 2457468.8507 | 0.5013 | -0.1487 | 2457468.8503 | 0.4995 | -0.2037 |
| 2457468.8528 | 0.5102 | -0.1451 | 2457468.8523 | 0.5079 | -0.1493 | 2457468.8519 | 0.5061 | -0.2023 |
| 2457468.8544 | 0.5169 | -0.1105 | 2457468.8538 | 0.5145 | -0.1589 | 2457468.8534 | 0.5127 | -0.2201 |
| 2457468.8560 | 0.5234 | -0.1153 | 2457468.8554 | 0.5211 | -0.1883 | 2457468.8550 | 0.5193 | -0.2244 |
| 2457468.8575 | 0.5301 | -0.1749 | 2457468.857 | 0.5277 | -0.1944 | 2457468.8565 | 0.5259 | -0.2382 |
| 2457468.8591 | 0.5366 | -0.2127 | 2457468.8585 | 0.5343 | -0.2392 | 2457468.8581 | 0.5325 | -0.2937 |
| 2457468.8607 | 0.5433 | -0.2403 | 2457468.8601 | 0.5409 | -0.2708 | 2457468.8597 | 0.5391 | -0.3016 |
| 2457468.8623 | 0.5499 | -0.2742 | 2457468.8617 | 0.5476 | -0.3262 | 2457468.8613 | 0.5457 | -0.3598 |
| 2457468.8638 | 0.5565 | -0.3326 | 2457468.8633 | 0.5542 | -0.3545 | 2457468.8628 | 0.5524 | -0.3887 |
| 2457468.8654 | 0.5632 | -0.3707 | 2457468.8649 | 0.5608 | -0.3898 | 2457468.8644 | 0.5590 | -0.4046 |
| 2457468.8670 | 0.5698 | -0.4800 | 2457468.8664 | 0.5674 | -0.3841 | 2457468.8660 | 0.5657 | -0.4706 |
| 2457468.8686 | 0.5765 | -0.4194 | 2457468.868 | 0.5741 | -0.4390 | 2457468.8676 | 0.5723 | -0.4922 |
| 2457468.8701 | 0.5830 | -0.4731 | 2457468.8696 | 0.5807 | -0.4720 | 2457468.8692 | 0.5789 | -0.5918 |
| 2457468.8717 | 0.5897 | -0.4907 | 2457468.8711 | 0.5873 | -0.5259 | 2457468.8707 | 0.5855 | -0.5699 |
| 2457468.8733 | 0.5963 | -0.5179 | 2457468.8727 | 0.5939 | -0.5549 | 2457468.8723 | 0.5921 | -0.5824 |
| 2457468.8748 | 0.6029 | -0.5462 | 2457468.8743 | 0.6005 | -0.5710 | 2457468.8739 | 0.5987 | -0.5938 |
| 2457468.8764 | 0.6095 | -0.5819 | 2457468.8759 | 0.6071 | -0.5771 | 2457468.8754 | 0.6054 | -0.6420 |
| 2457468.8780 | 0.6161 | -0.5868 | 2457468.8774 | 0.6137 | -0.6137 | 2457468.8770 | 0.6119 | -0.6417 |
| 2457468.8796 | 0.6227 | -0.6081 | 2457468.879 | 0.6203 | -0.6149 | 2457468.8786 | 0.6186 | -0.6609 |
| 2457468.8811 | 0.6293 | -0.6241 | 2457468.8806 | 0.627 | -0.6361 | 2457468.8801 | 0.6251 | -0.6684 |
| 2457468.8827 | 0.6359 | -0.6265 | 2457468.8821 | 0.6335 | -0.6272 | 2457468.8817 | 0.6318 | -0.6760 |
| 2457468.8843 | 0.6425 | -0.6117 | 2457468.8837 | 0.6402 | -0.6349 | 2457468.8833 | 0.6384 | -0.6999 |
| 2457468.8858 | 0.6491 | -0.6619 | 2457468.8853 | 0.6468 | -0.6452 | 2457468.8849 | 0.6450 | -0.6983 |
| 2457468.8874 | 0.6558 | -0.6577 | 2457468.8869 | 0.6534 | -0.6724 | 2457468.8864 | 0.6516 | -0.7134 |
| 2457468.8890 | 0.6623 | -0.6477 | 2457468.8884 | 0.66 | -0.6750 | 2457468.8880 | 0.6582 | -0.7104 |
| 2457468.8906 | 0.6690 | -0.6489 | 2457468.89 | 0.6666 | -0.6813 | 2457468.8896 | 0.6648 | -0.7183 |
| 2457468.8921 | 0.6756 | -0.6803 | 2457468.8916 | 0.6732 | -0.6954 | 2457468.8911 | 0.6715 | -0.7140 |
| 2457468.8937 | 0.6823 | -0.6623 | 2457468.8931 | 0.6799 | -0.6902 | 2457468.8927 | 0.6781 | -0.7236 |
| 2457468.8953 | 0.6889 | -0.7026 | 2457468.8947 | 0.6865 | -0.6820 | 2457468.8943 | 0.6848 | -0.7320 |
| 2457468.8969 | 0.6956 | -0.6847 | 2457468.8963 | 0.6932 | -0.7052 | 2457468.8959 | 0.6914 | -0.7408 |
| 2457468.8985 | 0.7022 | -0.7052 | 2457468.8979 | 0.6999 | -0.7000 | 2457468.8975 | 0.6980 | -0.7362 |
| 2457468.9000 | 0.7089 | -0.6862 | 2457468.8995 | 0.7065 | -0.7034 | 2457468.8990 | 0.7047 | -0.7608 |
| 2457468.9016 | 0.7155 | -0.6852 | 2457468.9011 | 0.7132 | -0.7086 | 2457468.9006 | 0.7114 | -0.7471 |
| 2457468.9032 | 0.7222 | -0.7036 | 2457468.9026 | 0.7198 | -0.7182 | 2457468.9022 | 0.7180 | -0.7620 |
| 2457468.9048 | 0.7289 | -0.7188 | 2457468.9042 | 0.7265 | -0.7100 | 2457468.9038 | 0.7247 | -0.7550 |

Table 2. New times of minima for the eclipsing binary J0923

| HJD | Error | Min | Filter |
|---|---|---|---|
| 2457468.7314 | 0.0002 | I | B |
| 2457468.7316 | 0.0001 | I | V |
| 2457468.7313 | 0.0002 | I | R |
| 2457468.8499 | 0.0002 | II | B |
| 2457468.8496 | 0.0001 | II | V |
| 2457468.8497 | 0.0001 | II | R |

The brightness difference between the two maxima (O'Connell) ($D_{max}$ = Max I − Max II) and minima ($D_{min}$ = Min I − Min II) for the observed light curves (LCs) have been measured and listed in Table 3 together with the depths of the primary ($A_p$ = Max I − Min I) and the secondary ($A_s$ = Max I − Min II) minima. The calculated values indicate the difference between the two maxima of the LCs in BVR bands, which often due to spots on the stellar surface (Lohr et al. 2012).

Table 3 Characteristic parameters of observed BVR light curves of the eclipsing binary J0923

| Parameter | B | V | R |
| --- | --- | --- | --- |
| $D_{max}$ (Max I – Max II) | 0.050±0.002 | 0.045±0.002 | 0.035±0.001 |
| $D_{min}$ (Min I – Min II) | -0.035±0.001 | -0.020±0.001 | 0.025±0.001 |
| $A_p$ = (Min I – Max I) | -0.450±0.018 | -0.415±0.017 | -0.365±0.015 |
| $A_s$ = (Min II – Max) | -0.365±0.015 | -0.350±0.014 | -0.355±0.015 |

## 4. Photometric Solution

The observed BVR light curves of the system J0923 were undergone to photometric analysis with the 2009 version of the Wilson and Devinney code (Nelson, 2009), which based on model atmospheres by Kuruz (1993). The temperature of the primary star ($T_1$) was estimated initially using the color index (B-V) = 0.987 corresponding to a spectral type about K3 from color index temperature relation by Tokunaga (2000). The initial temperature for star 1 was $T_1 = 4686\ ^0K$.

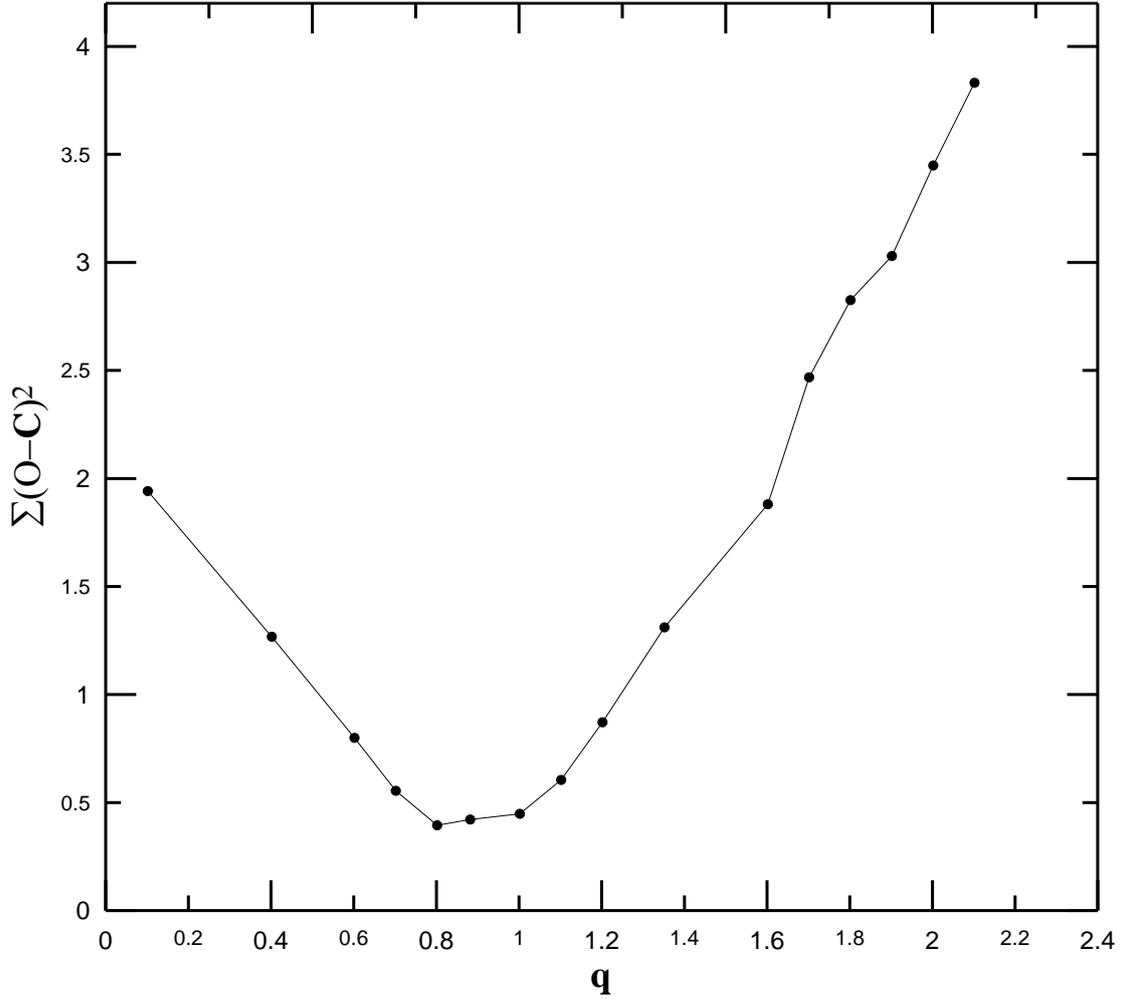

Figure 2. **q**-search for the system J0923.

Gravity darkening coefficients $g_1 = g_2 = 0.32$ (Lucy, 1967) and the bolometric albedo $A_1 = A_2 = 0.5$ (Rucinski, 1969) were adopted for convective envelopes ($T_{eff} < 7500\ ^0K$). The bolometric limb darkening coefficients ($x_1 = x_2$, $y_1 = y_2$) were adopted and interpolated using the square-root law from Van Hamme (1993). The commonly adjustable parameters through the light curve solution are The orbital inclination (i); the mean temperature of the secondary star ($T_2$); surface potential ($\Omega$), and the monochromatic luminosity of the primary star ($L_1$). The relative brightness of secondary star was calculated from stellar atmosphere model.

Since there is no any published radial velocity curve for the system J0923, there is no reliable mass ratio (q) can be used. A q-search was used to estimate the best initial value for the mass ratio (q) of J0923. Figure 2 represents the relation between the resulting sum of weighted square deviation $\Sigma(O-C)^2$ and (q). A minimum value of $\Sigma(O-C)^2$ is achieved at 0.8, which set as an initial value for the adjustable mass ratio. A differential correction was performed for (q) until all the free parameters converged and a set of solutions was derived. The final value of (q) for converged solutions listed in Table 4.

Mode 3 (overcontact) was applied with a synchronous rotation and circular orbit was assumed. some parameters kept fixed (i.e $T_1$, $g$, $A$, $x$). Firstly we tried to find a model solution without any spot (not shown here) which does not fit the observed light curves well, this may refer as mentioned earlier to the asymmetries in the observed light curves. As seen, the observed light curves show asymmetries, where the observed magnitude between phases 0.1 and 0.3 is brighter than that between 0.7 and 0.9, which could be explained by O'Connell effect. In order to construct theoretical light curves matching the observed ones, we suggest a hot spotted model which making the binary system appear brighter in the side it present. A spotted model was adopted with two hot spots; one on the primary and the other on the secondary which show a good matching between the observed and synthetic light curves. The parameters of the accepted model are listed in Table 4, which reveal that both the components of the system J0923 are of spectral type K3 (Popper, 1980) and the primary component is hotter than the secondary with about $\Delta T \approx 40^0 K$. Figure 3 displays the observed light curves together with the synthetic curves in BVR pass-bands.

Table 4. The photometric solution for J0923 in BVR bands.

| Parameter | B Filter | V Filter | R Filter | BVR |
|---|---|---|---|---|
| $\lambda$ | 4400 | 5500 | 7000 | _ |
| $i$ | 74.96 ± 0.20 | 74.85± 0.19 | 75.39 ± 0.16 | 74.79± 0.19 |
| $g_1 = g_2$ | 0.50 | 0.50 | 0.50 | 0.50 |
| $A_1 = A_2$ | 0.32 | 0.32 | 0.32 | 0.32 |
| $Q = (M_2 / M_1)$ | 0.8732 ± 0.0025 | 0.8757± 0.0026 | 0.8786 ± 0.0017 | 0.8752 ± 0.0023 |
| $\Omega$ | 3.5285 ± 0.0052 | 3.5229± 0.0062 | 3.5498±0.1449 | 3.5366 ± 0.0061 |
| $\Omega_{in}$ | 3.5408 | 3.5449 | 3.5498 | 3.5442 |
| $\Omega_{out}$ | 3.0558 | 3.0587 | 3.0623 | 3.0583 |
| $T_1$ $^0$K | 4690 Fixed | 4690 Fixed | 4690 Fixed | 4690 Fixed |
| $T_2$ $^0$K | 4635± 2 | 4641 ±2 | 4641± 2 | 4640± 2 |
| $L_1/(L_1+L_2)$ | 0.5477±0.0224 | 0.0526±0.0022 | 0.5394±0.0220 | _ |
| $L_2/(L_1+L_2)$ | 0.4523±0.0185 | 0.9475±0.0387 | 0.4606±0.0188 | _ |
| $r_1$ pole | 0.3679 ±0.0009 | 0.3702 ±0.0018 | 0.3670± 0.0031 | 0.3683± 0.0028 |
| $r_1$ side | 0.3872 ± 0.0011 | 0.3900 ±0.0023 | 0.3861 ± 0.0039 | 0.3877 ±0.0035 |
| $r_1$ back | 0.4179 ±0.0016 | 0.4218 ±0.0035 | 0.4166 ± 0.0058 | 0.4186± 0.0052 |
| $r_2$ pole | 0.3453 ±0.0009 | 0.3480 ±0.0019 | 0.3454 ± 0.0033 | 0.3461 ±0.0030 |
| $r_2$ side | 0.3622 ± 0.0011 | 0.3654 ±0.0024 | 0.3622 ± 0.0041 | 0.3631± 0.0037 |
| $r_2$ back | 0.3938 ± 0.0017 | 0.3983 ±0.0037 | 0.3936 ± 0.0061 | 0.3949± 0.0056 |
| *Spot parameters for star 1* | | | | |
| Co-latitude (deg) | 100 Fixed | 100 Fixed | 100 Fixed | 100 Fixed |
| Longitude (deg) | 120 Fixed | 120 Fixed | 120 Fixed | 120 Fixed |
| Spot radius (deg) | 23.630±0.465 | 22.500±0.919 | 27±1.10 | 22.29±0.737 |
| Temp. factor | 1.099±0.002 | 1.080±0.044 | 1.060±0.043 | 1.089±0.003 |
| *Spot parameters for star 2* | | | | |
| Co-latitude (deg) | 140 Fixed | 140 Fixed | 140 Fixed | 140 Fixed |
| Longitude (deg) | 80 Fixed | 75 Fixed | 85 Fixed | 75 Fixed |
| Spot radius (deg) | 27±1.102 | 27±1.102 | 29±1.184 | 27±0.131 |
| Temp. factor | 1.333±0.004 | 1.330±0.054 | 1.330±0.05 | 1.31±0.005 |
| $\sum (O-C)^2$ | 0.00736 | 0.00786 | 0.00640 | 0.08279 |

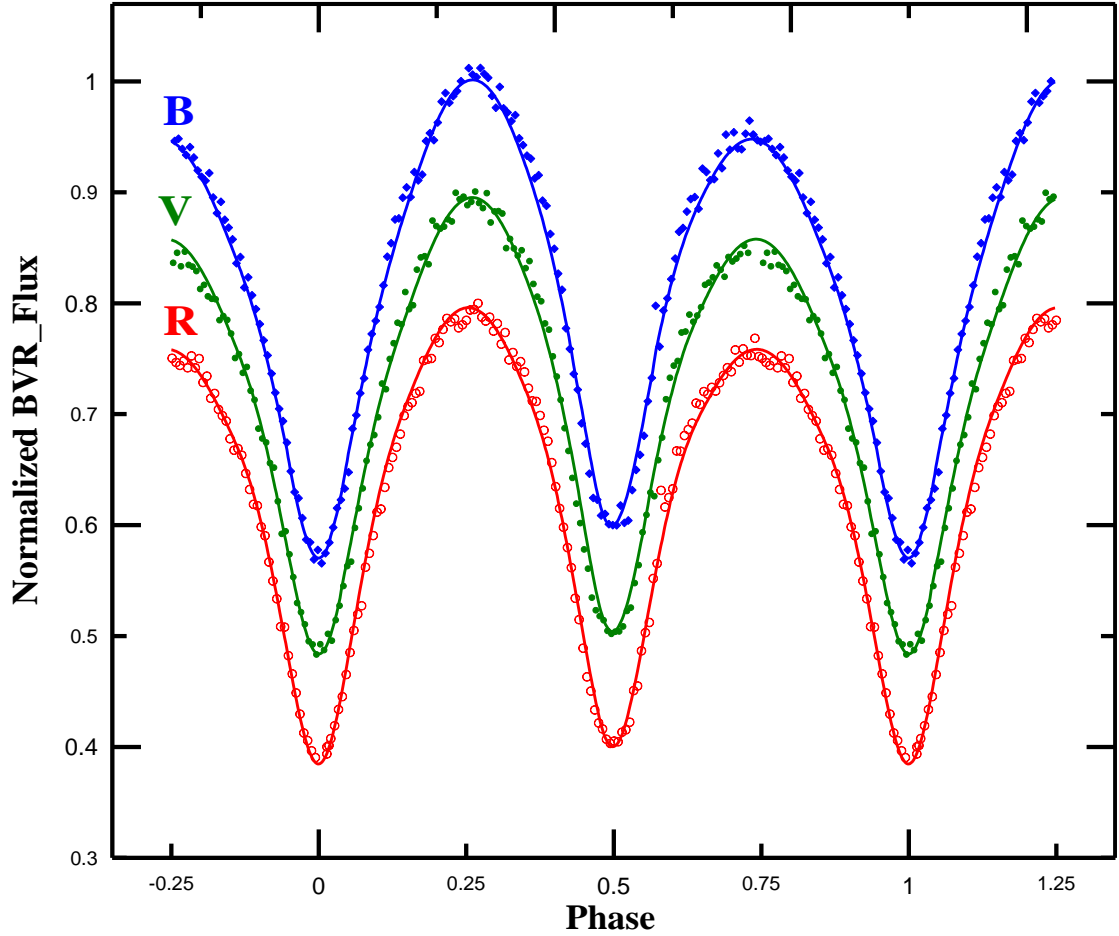

Fig. 3 Observed light curves (*filled circles*) and synthetic light curves (*solid lines*).

Table 5 Absolute physical parameters for the system J0923

| Element | M(M$_\odot$) | R(R$_\odot$) | T(T$_\odot$) | L(L$_\odot$) | M$_{bol}$ | Sp. type | Log g (cgs) |
|---|---|---|---|---|---|---|---|
| Primary | 0.73 ±0.03 | 0.82±0.03 | 0.81±0.03 | 0.29±0.01 | 6.10±0.25 | K3 | 4.52 |
| Secondary | 0.64±0.03 | 0.81±0.03 | 0.80±0.03 | 0.27±0.01 | 6.17±0.25 | K3 | 4.52 |

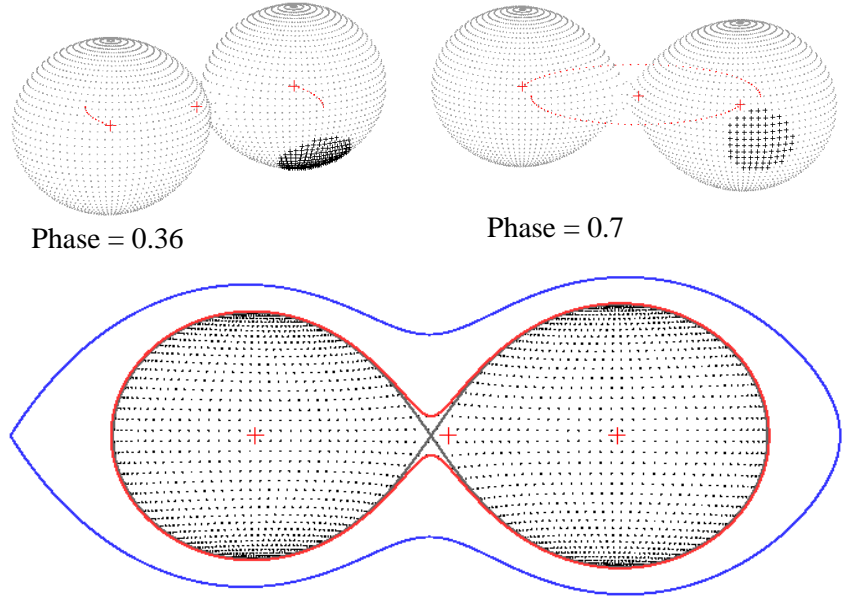

Fig 4. Geometric structure of the binary system J0923.

The absolute physical parameters for both components of the system J0923 were calculated using the empirical $T_{eff}$ − Mass relation by Harmanec (1988), (see Table 5). The estimated mass of the primary component is $M_1 = 0.7303 \pm 0.0298\ M_\odot$, while the mass of the secondary component is $M_2 = 0.6392 \pm 0.0261\ M_\odot$. Radii of the system components $R_1(R_\odot)$, $R_2(R_\odot)$ and bolometric magnitudes ($M_{bol}$) were calculated. It's clear that the accepted photometric solution and the estimated physical parameters show that the primary component is hotter and more massive than the secondary one. Three-dimensional geometrical structure for the system J0923 was constructed as shown in Figure 4 using the software Package Binary Maker 3.03 (Bradstreet and Steelman, 2004) based on the calculated parameters resulting from our model.

## 5. Discussion and Conclusions

New CCD observations for the newly discovered W UMa system J0923 were carried out in BVR filters. A total of 8 new timings of minima (4 primary and 4 secondaries) were calculated in each filter using Kwee and Van Worden (1956) method. Period of the system was upgraded according to our observations and the first ephemeris was adopted using the new minima and new period.

Light curve modeling was performed using the complete light curves by means of the W-D code shows that the system J0923 is an overcontact binary. The asymmetric light curves can be explained by a spotted model with two hot spots. The orbital elements were estimated, which reveals that the primary star is more massive and hotter with about $\Delta T \approx 40^0 K$ than the secondary one.

According to the adopted temperatures of the primary star ($T_1$), and the secondary one ($T_2$), both components are belonging to spectral type K3 (Popper, 1980). The absolute physical parameters were calculated and used to identify and follow the evolutionary for the components of the studied system.

We use the preliminary physical parameters of the components of the system J0923 listed in Table 5 to investigate the evolution of the system. For this purpose, we used mass-luminosity (M-L) and mass-radius (M-R) relations for zero-age main sequence (ZAMS) and thermal age main sequence (TAMS) with metalicity $z = 0.019$, Girardi et al. (2000). Figures 5 and 6 display the M-L and M-R relations for the components of the system J0923. From the figures, the primary and the secondary components are located above the ZAMS, which indicates that the two components are an evolved star.

In Figure 7, we compared our physical parameters with the mass-Teff relation for intermediate and low mass stars based on data of detached double-lined eclipsing binaries by Malkov (2007). The locations of the two components display the same behaviour as M-L and M-R relations with a slight deviation of the secondary component. More photometric and spectroscopic observations for the system J0923 are recommended to follow and study its period behavior, mass ratio and evolution of the system.

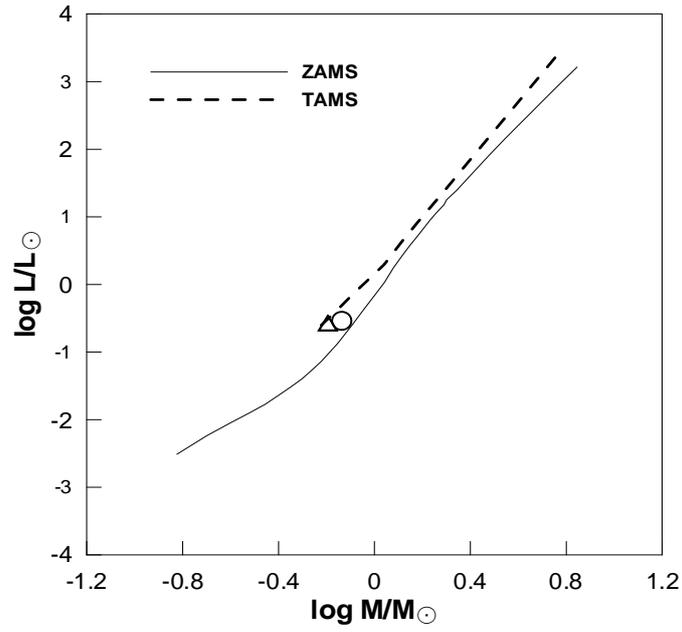

Fig. 5: Position of the components of the system J0923 on the Mass-Luminosity relation Girardi et al. (2000).

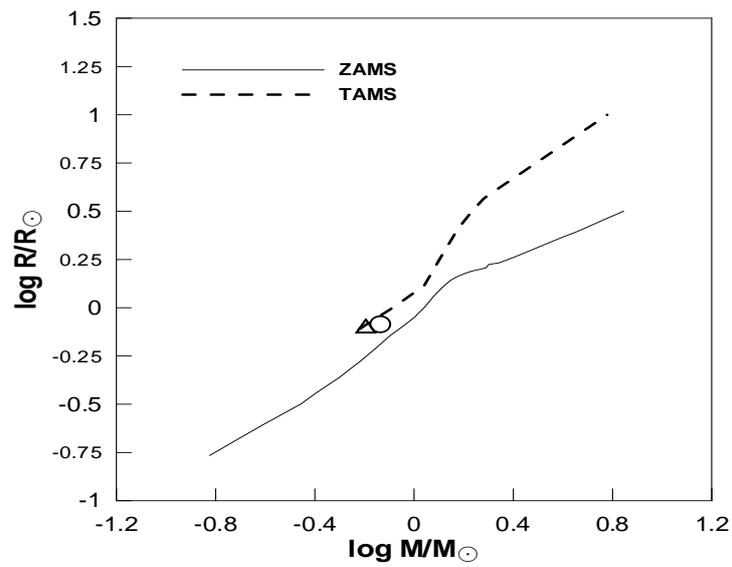

Fig. 6: Position of the components of the system J0923 on the Mass-Radius relation Girardi et al. (2000).

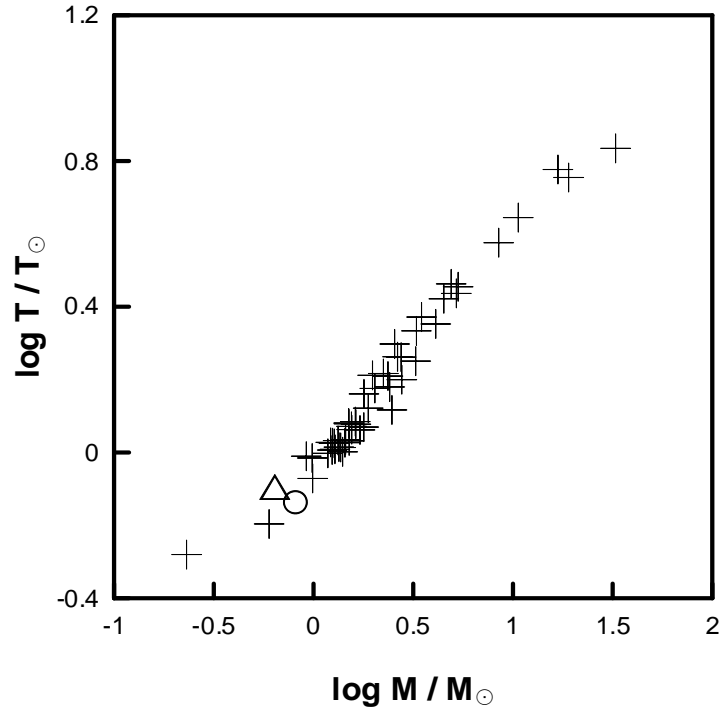

Fig. 7: Position of the components of the system J0923 on the Mass - Temperature relation (Malkov, 2007).